\documentclass[pre,superscriptaddress,12pt]{revtex4-1}
\usepackage{graphicx}
\begin{document}

\title{Organizational structure and communication networks in a university environment}
\author{Joachim Mathiesen$^{1,2}$, Bj\o rn Jamtveit$^2$, and Kim Sneppen}
\affiliation{Niels Bohr Institute, University of Copenhagen, Blegdamsvej 17, DK-2100 Copenhagen, Denmark\\ $^2$Physics of Geological Processes, University of Oslo, P.O. Box 1048 Blindern, N-0316 Oslo, Norway}
\begin{abstract} 
The ``six degrees of separation" between any two individuals on Earth has become emblematic of the 'small world' theme, even though the information conveyed via a chain of human encounters decays very rapidly with increasing chain length, and diffusion of information via this process may be very inefficient in large human organizations. The information flow on a communication network in a large organization, the University of Oslo, has been studied by analyzing e-mail records. The records allow for quantification of communication intensity across organizational levels and between organizational units (referred to as ``modules"). We find that the number of e-mails messages within modules scales with module size to the power of $1.29\pm .06$, and the frequency of communication between individuals decays exponentially with the number of links required upwards in the organizational hierarchy before they are connected. Our data also indicates that the number of messages sent by administrative units is proportional to the number of individuals at lower levels in the administrative hierarchy, and the ``divergence of information" within modules is associated with this linear relationship. The observed scaling is consistent with a hierarchical system in which individuals far apart in the organization interact little with each other and receive a disproportionate number of messages from higher levels in the administrative hierarchy.
\end{abstract}

\maketitle

\section{Introduction}
Network studies have, to a large extent, focused on nearest neighbor interactions \cite{M67,AB02}, implicitly assuming that the mere existence of a connection implies that transmission of information is complete or of equal quality for all connections. This connotation is associated with the fact that most studies focus only on the structure imposed by the network itself \cite{N06,BLMCH06,STR05}. However a given network typically represents only part of a much larger system that is connected in a variety of ways. In the case of social systems, for example, the network is embedded in a physical world where a single email exchange between two persons is not enough to establish a significant or lasting link. In a wider context, the six degrees of separation paradigm, based on the vague concept of two individuals ``knowing"' each other, is not a realistic measure of the efficiency of information exchange \cite{F82}. On the contrary, if individuals communicated with everybody on Earth via a chain of human encounters, the average information conveyed by any particular chain would necessarily have to be vanishingly small.

Here we consider how the intensity of e-mail exchange decreases as function of distance defined by the number of links required upwards in the modular hierarchical organization to connect the sender and recipient. As an example, we studied the internal e-mail communication between all 5600 employees and 30,000 students at the University of Oslo. More than 3.6 million internal email messages between more than 30,000 active email accounts during one month (April 2009) were analyzed. The communications were mapped onto a directed network by identifying individual email accounts as nodes and email messages as directed links. In Sec. II, the intensity of email communication between individual units in the organization is considered and in Sec. III, we propose a model of the organizational hierarchy. Finally, Sec. IV offers a discussion. 

\begin{figure}
\includegraphics[width=.5\textwidth]{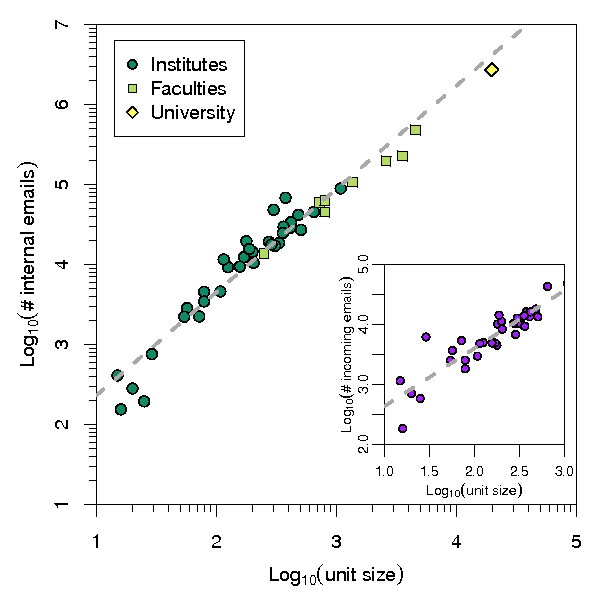}
\caption{(Color online) Scaling of the intensity of communication within individual departments. The number of messages versus the number of active email accounts shown on a double-logarithmic scale. The dashed line is a best fit with a slope of $1.29\pm 0.06$ s.d. The inset shows the number of incoming messages from individuals external to the departments (the dashed line is a best fit with a slope $1.0\pm 0.1$ s.d.)}\label{scaling}
\end{figure}%
\section{Communication network}
While the topology of email networks has previously been studied \cite{EMB02,GDDGA03,NFB02,EMS04} with a focus on degree distributions, clustering coefficient, temporal structures and community distribution \cite{TWH05}, we investigated the e-mail exchange in an environment where the classification in terms of organizational (departmental) modules is used to analyze the communication network. The network exhibits broad  distributions of the in- and out-degree $k$, $n_{in/out}(k)\sim k^{\delta_{in/out}}$, with exponents $\delta_{in}=-1.9$ and $\delta_{out}=-1.2$ reproducing previously reported findings for e-mail networks \cite{EMB02}. More interestingly, Fig.~1 shows the number of messages between people belonging to the same unit versus unit size (here a unit is a department, faculty or the full university). This internal communication is consistent with a power-law relating the number of messages to unit size with an exponent of $\gamma=1.29\pm 0.06$ s.d. For comparison, in a scenario where each individual communicated with a fixed number of individuals, the exponent would be $\gamma=1.0$, whereas an exponent of $\gamma=2.0$ would be expected if everyone interacts with everyone else within a unit.

Moreover, the inset of Fig.~1 indicates that the number of external messages originating from a department increases linearly ($\gamma = 1.0\pm 0.1$ s.d.) with department sizes. These scaling laws do not change if communications sent simultaneously to more than a few individuals are removed.
While individuals in larger departments are not expected to send more messages to external addresses or to send more messages in total, they do communicate more with people in their own department. Overall the super linear growth of communication within sub-units indicates that larger groups devote disproportionally more resources to internal information flow.  In practice, individuals in departments with 550 members (including students) send twice the number of internal messages compared with individuals in departments of size 100.

\begin{figure}
\includegraphics[width=.5\textwidth]{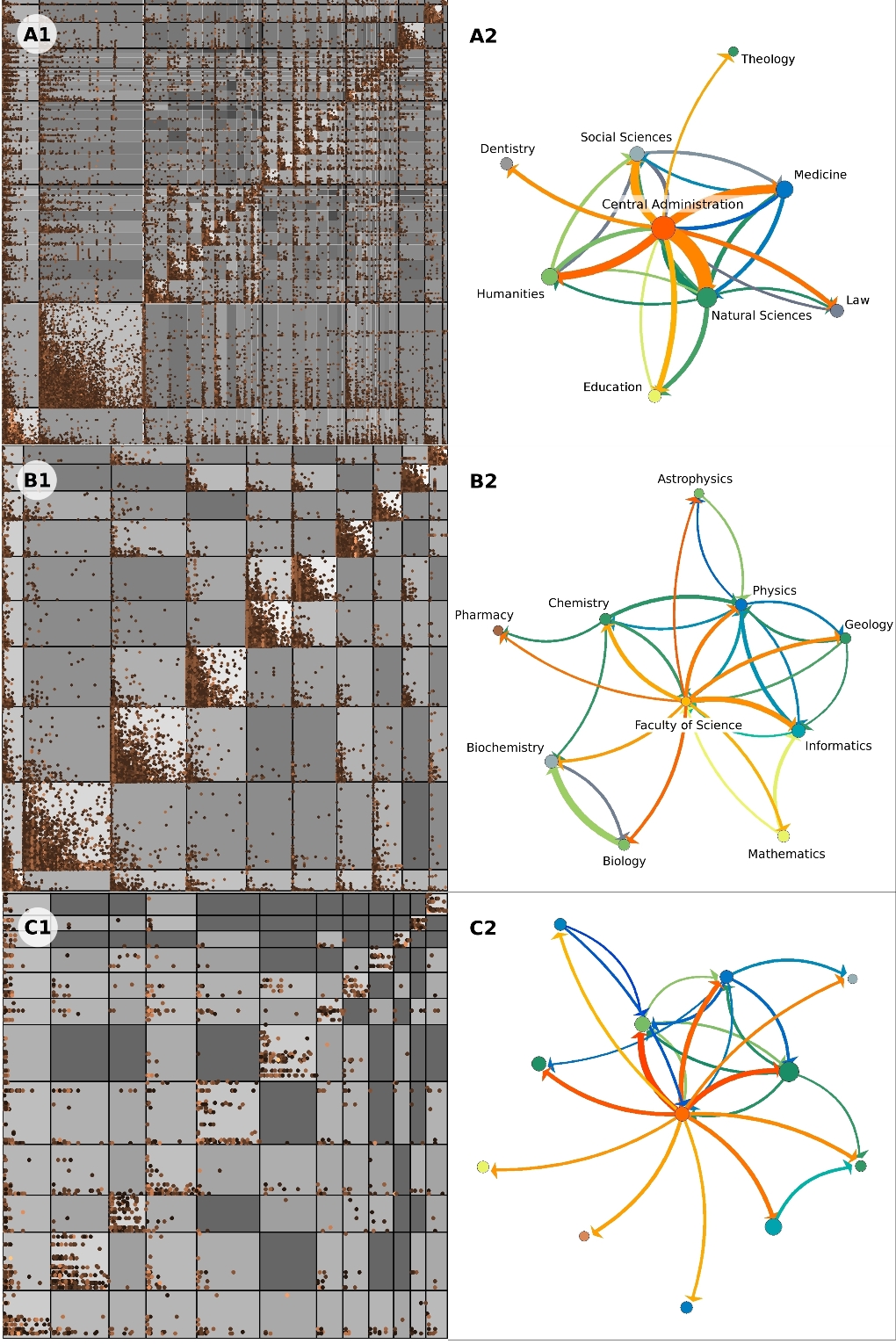}%
\caption{(Color online) Directed email communication networks at different levels of the organization. Left panels show adjacency matrices for the communication network a) between faculties and administration, c) within the Faculty of Mathematics and Natural Sciences and e) within the Department of Physics. The background color scale indicates the logarithm of the number of messages between various units. The panels in the right-hand-side b), d) and f) show corresponding network diagrams for the information flow between units. The width of a link is proportional to the square root of the number of messages sent (in the lower right panel the width is increased by a factor of 2 compared to the upper panels). The central administration sends out 2.5-5.2 times as many messages to other units as it receives. The same number in panel d) is 3.5 (1.5-8) and 7.0 (1.25-17) for the central hub in f). Links with intensities smaller than 10\% of the maximum intensity are not shown in the right-hand panels.  
}\label{networks}
\end{figure} 
\section{Hierarchical organization}
The University of Oslo has a 3-tier organization with a central administration including several support units, 8 faculties and a number of departments. Fig.~2a shows a matrix of all messages between individuals at the university where each dot corresponds to messages send from a user on the x-axis to a user on the y-axis. If a grid element has a lighter color, it means that more emails are being sent between pairs of users within ``the grid element". On each axis, the individuals in the network are sorted first according to major units e.g. administration or faculty, and each of these groups is then sorted according to department affiliation. Finally within all these groups, individuals are sorted according to email activity. Immediately it becomes clear that the network has a strong modularity which primarily reflects boundaries between disciplines. A modularity that is further emphasized by panel B and C highlighting the modular structures of
the Faculty of Science, and of the Department of Physics. The Department of Physics was divided into modules using an information-theoretic approach \cite{RB07}. The right hand panels of Fig.~2 show the corresponding networks where the thicknesses of the links indicate the intensity of communication. In general the communication is strongly directed at all levels, the central hubs  (administrations at various levels) sends out many more emails to their sub-ordinates than they receive back. 
\begin{figure}
\includegraphics[width=.5\textwidth]{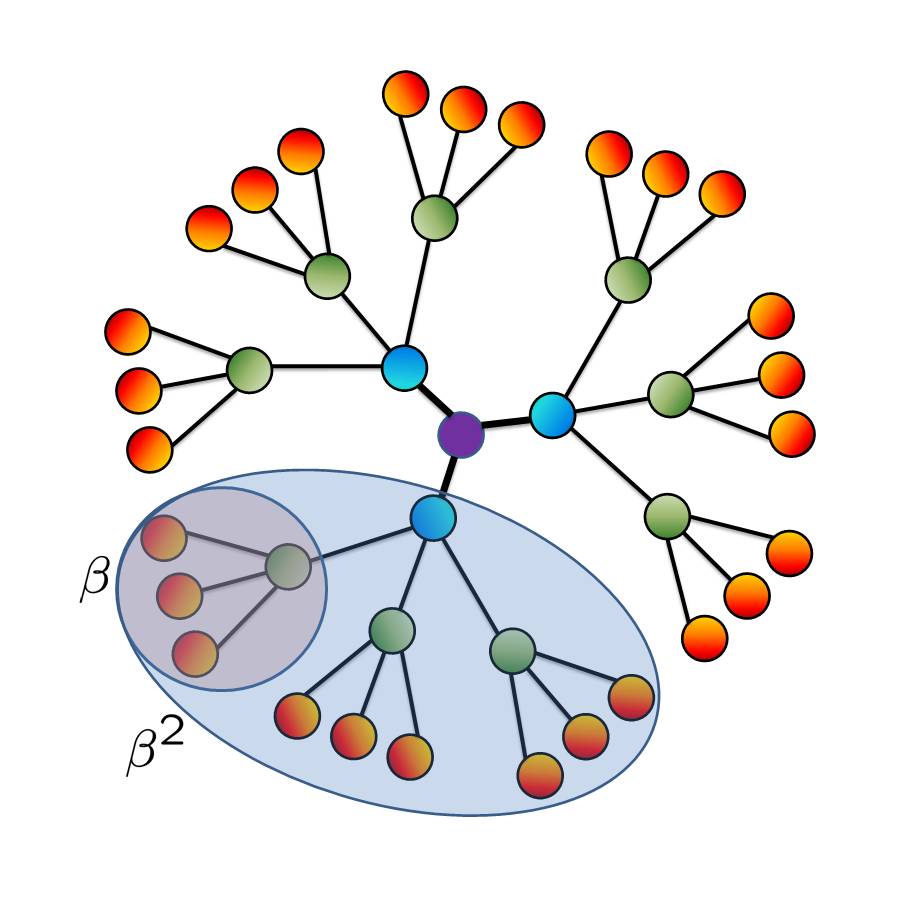}
\caption{(Color online) Simple model of the organizational structure and information flow on the communication network.}\label{setup}
\end{figure}

To understand how the super linear scaling appears from the apparent self-similar structure of the networks in Fig.~2, we propose a minimal model of the communication network based on a simple hierarchy with one unit on top (the central administration, level $0$) and with each unit at level $k$ connected to $B$ lower ranked (level $k+1$) units. The total number of messages $E$ is 
then expressed as a sum over the organizational levels
\begin{equation}\label{sum}
E=\sum_{k=0}^n \alpha_k B^k.
\end{equation}
Here $n$ is the number of levels in the hierarchy and $\alpha_k$ is the total
number of messages that a unit at level $k$ receives from units at higher levels or 
at same level in the hierarchy (See Fig.~3).
It is assumed that the number of messages that an individual receives from someone else is proportional to $\beta^{|\delta k|}$, where $\delta k$ is the number of levels they have to go up until they both belong to the same unit. In this way the expression 
\begin{eqnarray}\label{alpha}
\alpha_k &=&C_0\left[1+\beta B+\ldots+\beta^k B^{k}\right]\nonumber\\&=&C_0\frac{1-\beta^{k+1}B^{k+1}}{1-\beta B}
\end{eqnarray}
is established for $\alpha_k$, where $C_0$ is a measure of the number of messages generated in a single unit.
By inserting this expression in Eq. (\ref{sum}) and by assuming that $\beta$ is sufficiently larger than $1/B$ we achieve the approximation \begin{equation}
E\sim \beta^n B ^{2n}=N^{2-\frac{\log 1/\beta}{\log B}}
\end{equation}
for the total number of messages sent.
Here we have assumed that the number of levels $n$ in the hierarchy is related to the number $N$ of people in the organization by $n=\log N/\log B$. The reduction factor $\beta$ is estimated from the email data by measuring the ratio $\kappa$ between the number of messages between a single research group in Fig.~3 and their department and the number of messages between the same group and the whole faculty. This ratio is estimated to be $\kappa = .45\pm .08$. Finally using Eq. (\ref{alpha}), we arrive at the following estimate $\beta=\frac 1 {\kappa B}$
Assuming that $B\approx 10$ we arrive at the scaling law $E\sim N^{1.3}$, which is consistent with the observed power law in Fig.~1. The communication intensity is estimated to decrease by a factor $\beta=0.22$ for each level in hierarchy that two individuals have to go up before they belong to the same unit. This may be illustrated in the following way. If a typical communication between two physicists is 10 messages, there should be 2.2 emails between a physicist and a mathematician, whereas there would only be $\beta \times 2.2=0.22\times 2.2\approx 0.5$ emails between a physicist and a linguist. However, this is not true, since most communication turns out to be dominated by nodes directly upstream in the hierarchy. For example within the faculty of Mathematics and Natural Sciences, as many as 44\% of the total messages outside departments is with the management of the faculty. Thus, roughly half of the communication is in relation to administrative nodes in the network. In the context of our scaling parameters, the number of messages from an individual to everyone at a hierarchical distance $k$ would scale as $\alpha_k  \sim (\beta B)^k\sim 2^k$, implying an increased intensity of emails with increased distance. Half of $B\beta$ is associated with administration, and the exponential increase would be marginal were it not for upstream communication. 

\section{Discussion}
In general, the structure of complex social organizations, which imposes constraints on the information flow that guides the life in the organization is a product of a long evolutionary process. One may ask whether the organization is, in some sense, optimal, for example in terms of cost-effectiveness. In this regard it is interesting that the anomalous scaling of communication with system size (Fig.~1) and the scaling of support staff with academic staff reported in \cite{JJM09} are similar. In our case, the divergence in communication is tightly coupled to increased vertical communication with administrative units. The anomalous scaling in \cite{JJM09} is associated to a larger administration (relative to total number of individuals) for larger sub-systems. In both cases the data support a hierarchical model in which the academic nodes are almost invariably located at the bottom of the hierarchy, they are strongly modulated (disconnected) and they are connected mostly through administrative units. Similar hierarchical structures have also been observed in the social networks of open source communities \cite{VS07}.
Another issue is to what extent hierarchical organizations, subdivided according to disciplines, are efficient \cite{PGAMS99} and might inhibit cross-disciplinary activities. Hierarchical organizations appear to be optimal for top-down information flow whereas they often become inefficient for horizontal interactions. An overload of vertical communication at the level of the ``primary producers" could seriously weaken an organization and limit the ability to generate synergy effects from horizontal communication. In interpreting our data, it is important to take into account communication outside the organization. For a university, communication with the outside world, justifies the organization in a broad sense. It goes without saying that the organization structure should ensure that non-essential interior communication is kept minimal in order to leave resources for impacting outside society. From this perspective our observations favor several smaller independent organizations instead of one large one.
\begin{acknowledgments}
Suggestions and comments by Paul Meakin are gratefully acknowledged. This study was supported by a Center of Excellence grant to the Physics of Geological Processes Centre at the University of Oslo and by the Danish National Research Foundation through the Center for Models of Life.
\end{acknowledgments}

\end{document}